\documentstyle[aps]{revtex}
\title{The pseudogap in underdoped high $T_c$ superconductors 
       in the framework of the Boson Fermion model}
\author{J.~Ranninger and J.-M.Robin} 
\address{Centre de
Recherches sur les Tr\`es Basses Temp\'eratures, Laboratoire
Associ\'e \'a l'Universit\'e Joseph Fourier, Centre National de la
Recherche Scientifique,\\ BP 166, 38042, Grenoble C\'edex 9, France}
\date{\today} 
\begin{document} 
\maketitle 
\draft 
\begin{abstract}

The question of whether the pseudogap in high $T_c$ cuprates is 
related to superconducting precursor effects or to the existence 
of extrinsic bosonic massive
excitations is investigated on the basis of the Boson-Fermion model.
The characteristic three peak structure of the electronic spectral
function and the temperature dependent Fermi vector derived here are  
signatures for a two component scenario which can be tested by ARPES 
and BIS experiments.


\end{abstract} 

%\narrowtext 

%\twocolumn

\newpage 
\section{INTRODUCTION}

The recent direct experimental verifications\cite{Marshall-96,Ding-96} 
of a pseudogap in the density of states of the electrons in the metallic 
layers of the cuprate high temperature superconductors in the underdoped 
regime is presently considered to be one of the key 
elements to guide us in unravelling the underlying mechanism for 
superconductivity in these materials.

The most immediate temptation to explain the pseudogap which opens at a 
certain temperature $T^{*}$, sometimes well above the superconducting 
transition temperature, is to consider it as a consequence of 
superconducting phase fluctuations corresponding to a finite and relatively 
long lived 
amplitude of the order parameter above $T_c$. Support in favor of such 
a scenario comes from the fact that the symmetry of the pseudogap is 
the same as that of the true gap and that the materials are quasi 2D 
structures, for which Cooper pairs are expected to exist in form of 
propagating states above $T_c$\cite{Hohenberg-67}. 
It is in this context that many theoretical 
papers base their study on the intricate interplay between electronic 
quasi-particles and their bosonic resonant states.

Scenarios for the pseudogap based on systems with 
attractive interaction between the electrons exhibit such a precursor 
pseudogap  in the cross-over regime between BCS an a superfluid state 
of preformed 
tighly bound electron pairs. 
Quite generally and independent of any particular scenario $T_c$  
is then determined by a temperature  $T_{\Theta}$ characterizing the onset 
of phase fluctuations of the order parameter, while the meanfield critical 
temperature $T_{MF}$ describes the formation of pairs. 
This  leads to a suppression of intensity of low lying single particle 
excitations 
and thus to the appearance of a pseudogap\cite{Emery-95}. 
Specific scenarios attempting to describe such behavior are based on generic 
models for superconductivity with some effective attractive 
interaction\cite{Maly-96,Janko-97,Blaer-97} and on the negative 
U Hubbard problem\cite{Micnas-95,Randeria-95}.

Contrary to such scenarios implicating a direct precursor effect, others
have been proposed where the pseudogap is induced by antiferromagnetic 
spin fluctuations rather than pair fluctuations. 
For single component systems such scenarios are based on strong repulsive 
interactions between the electrons. 
The pseudogap is then related to a SDW and/or Hubbard pseudogap close to 
a corresponding 
transition\cite{Schrieffer-95,Chubukov-95}. 
More precicely, the pseudogap features in the charge sector are derived 
from the pseudogap features of bosonic resonant states in the spin 
sector\cite{Shen-97,Jaklic-97,Preuss-97}.
A somewhat different scenario for the pseudogap is presented for the RVB 
physics, according to which chargeless spinons pair up at $T^{*}$ 
and holons condense at $T_c$\cite{Anderson-97}. Similar in spirit are the
$SU(2)$ symmetry meanfield studies of slave boson approaches in strongly 
correlated systems\cite{Wen-96}, which relate the pseudogap to the transition 
between small and large Fermi surfaces. 
Van Hove singularities also may play a role and were shown to  stabilize 
d wave RVB states with  pseudogaps near those singularities\cite{Ioffe-96}.

Apart from those single component scenarios, scenarios involving 
two component systems have been proposed.
Experiments on the optical conductivity have for a long time suggested that 
in high $T_c$ cuprates two types of carriers may be at stake, almost 
localized ones 
and itinerant ones\cite{Mihailovic-97}. 
Recent experiments on the specific 
heat are in support of such a hypothesis\cite {Loram-97} and chemical 
analysis\cite{Roehler-96} favors the picture of metallic layers imbedded 
in highly polarizable dielectrics which could be the seat of localized polarons 
or bi-polarons. These results clearly point in the direction of two component 
scenarios such as described by 
the generic Boson-Fermion (B-F) model introduced by us along time 
ago\cite{Ranninger-85}. On the basis of this model we have initially shown 
the opening of a pseudogap in the density of states of the Fermionic subsystem
setting in below a certain temperature $T^{*}$, the breakdown of Fermi liquid 
properties\cite{Ranninger-95,Ranninger-96} and the manifestations of that
in the thermodynamic, magnetic and transport 
properties\cite{Ranninger1-96,Ranninger-97}.
A scenario related to that has been proposed on the basis of stripes 
in the metallic layers, where 1D spin stripes alternate with 1D metallic 
stripes\cite{Emery-97}. The spin
stripes produce bosonic entities in form of tightly bound spin singlets which 
are exchanged by pair hopping processes with electron pairs in the 
neighboring metallic stripes.

The purpose of this communication is to elucidate the nature of the pseudogap 
as given in the scenario of the Boson-Fermion model. 
Is it due essentially to a precursor effect in low dimensional 
systems\cite{Ren-97} or is related to a two particle resonant state 
triggered by the existence of extrinsic 
(in contrast to intrinsic in single component systems) bosonic
modes such as localized bi-polarons which form part of the charge carriers 
of the system?
For this purpose we shall try to interpret our previous numerical 
calculations\cite{Ranninger-95,Ranninger-96} in terms of an approximative 
analytic approach. This will permit us to intuitively understand the three 
peak structure for the spectral properties of the electronic 
subsystem, the reasons for the breakdown of Fermi liquid properties and 
indicate to what extent the opening of the pseudogap 
is  related to a precursor to superconductivity. 
Moreover we shall show that for this model the Fermi vector will change 
as a function of temperature.
This should in principle be measureable and thus help to clarify whether 
high $T_c$ superconductors are one or two component systems.
 
\section{An analytic approach to understand the pseudogap}

The B-F  model describes a system of localized Bosons hybridized with 
conduction electrons. The Bosons can be thought of as bipolarons 
(localized electron pairs with opposite spin) mainly located 
in the dielectric layers  
surrounding the $CuO_2$ planes which contain the Fermions 
(conduction electrons)\cite{Ranninger1-95}. 
The corresponding Hamiltonian for 
this model is given by
 
\begin{eqnarray}
H & = & (zt-\mu)\sum_{i,\sigma}c^+_{i\sigma}c_{i\sigma}
-t\sum_{\langle i\neq j\rangle,\sigma}c^+_{i\sigma}c_{j\sigma}
+(\Delta_B-2\mu) \nonumber 
\\& & \times \sum_ib^+_ib_i
+v\sum_i [b^+_ic_{i\downarrow}c_{i\uparrow}
+c^+_{i\uparrow}c^+_{i\downarrow}b_i] 
\label{eq1}
\end{eqnarray}
where $c_{i\sigma}^{(+)}$ refers to the Fermion operators of the 
itinerant electrons and $b^{(+)}_i$  to the Boson operators.
The spin indices are given by $\sigma$, 
the bare hopping integral for 
the electrons is given by $t$, the Boson energy level by $\Delta_B$ 
and the Boson-Fermion pair exchange coupling constant by $v$.
$i$ denotes some effective site involving 
adjacent molecular clusters of the metallic and 
dielectric planes, which, as far as the metallic 
planes are concerned, should involve a unit of two neighboring 
$CuO_4$ clusters and as far as the dielectric planes are concerned 
could be thought of two adjacent apex oxygens together with their 
nearby dopant atoms. Such pairs of electrons, corresponding to the term
$c^+_{i\uparrow}c^+_{i\downarrow}$
in the metallic layers should be considered as intersite rather than 
onsite pairs.
Correlation effects which are certainly important are totally neglected here, 
because we want to focus on physics introcuced by pair hopping between 
the metallic and the dielectric layers.
Since we are dealing here with a two-component system we have to 
impose a common chemical potential $\mu$ 
in order to guarantee charge conservation.

We have previously evaluated the Fermion spectral functions 
and the density of states for this model for a one and respectively 
two dimensional bare (i.e., 
in the absence of Boson Fermion hybridization, $v=0$) density of states 
using a selfconsistent lowest order fully conserving diagramatic 
procedure\cite{Ranninger-95,Ranninger-96}. 
In the present work we  focuse on the normal state temperature 
variation of the 
spectral function for the Fermions at wave vector ${\bf k}_F$ as well as 
on the imaginary part of the self energy as a 
function of frequency. Since we are only interested here in 
qualitative features such as the temperature variation of the pseudogap,
we carry out our study for the case of a 1D bare density of states,
knowing from past experience\cite{Ranninger-95,Ranninger-96} that
the dimensionality only quantitavely influences the pseudogap features 
in this model. 
We therefore adopt the same procedure as outlined in 
ref\cite{Ranninger-95} but with a much denser mesh 
of 1000 ${\bf k}$ vectors in the Brillouin zone. 
This is required in order to track the temperature 
effects on the 
spectral function and to dertermine ${\bf k}_F$ 
as a function of temperature, 
given the fact that for our two component system the number 
of Fermions and Bosons is not conserved separately.

Before reporting on these numerical results, 
let us consider some basic features of the B-F model from which the 
underlying physics can be undestood.
The self energies for the Fermions $\Sigma_F({\bf k},i\omega_n)$ and 
Bosons  $\Sigma_B({\bf q},i\omega_m)$  within lowest order perturbation 
theory are given by
\begin{equation}
\Sigma_F({\bf k},i\omega_n)=\frac{v^2}{N}\sum_{\bf q}
{{n_F(\varepsilon_{-{\bf k}+{\bf q}})+n_B(E_0)}
\over {i\omega_n -E_0 + \varepsilon_{-{\bf k}+{\bf q}}}}
\end{equation}

\begin{equation}
\Sigma_B({\bf q},i\omega_m)=\frac{v^2}{N}\sum_{\bf k}
{{1-n_F(\varepsilon_{-{\bf k}+{\bf q}})-n_F(\varepsilon_{\bf k})}
\over {i\omega_m - \varepsilon_{-{\bf k}+{\bf q}}- \varepsilon_{\bf k}}}
\end{equation}
where $n_{F,B}(...)$ denote the Fermi and respectively 
Bose distribution function. 
$\varepsilon_{\bf k} \equiv t(z-\sum_{\bbox \delta}
exp(i{\bf k}{\bbox \delta})-\mu)$ is
the free Fermion dispersion with $\bbox \delta$ being the vectors 
linking nearest neighbour sites, $E_0 \equiv \Delta_B - 2 \mu$ and $N$ 
is the 
number of ${\bf k}$ vectors in the Brillouin zone.
Considering  the problem of the B-F model within the framework of a 
selfconsistent RPA scheme requires to look for distribution functions 
$n^F_{\bf k}$ and  $n^B_{\bf q}$ which replace $n_F(...)$ 
and $n_B(...)$ in Eqs.(2,3) and which have to be determined 
selfconsistently via the equations

\begin{equation}
n^F_{\bf k} = \frac{1}{2} 
+ \frac{1}{\beta}\sum_{i\omega_n}G_F({\bf k},i\omega_n) \;\;\;
n^B_{\bf q} = -\frac{1}{2} 
- \frac{1}{\beta}\sum_{i\omega_m}G_B({\bf q},i\omega_m)
\end{equation} 
with the Fermi and Bose Green's functions given by

\begin{eqnarray}
G_F({\bf k},i\omega_n)=[i\omega_n-\epsilon_{\bf k}
-\Sigma_F({\bf k},i\omega_n)]^{-1}, \\
G_B({\bf q},i\omega_m)=[i\omega_m-E_0-\Sigma_B({\bf q},i\omega_m)]^{-1}
\end{eqnarray}

Contrary to the conserving diagramatic approach to this problem  
which we have solved previously\cite{Ranninger-95,Ranninger-96}, 
we are unable to find 
numerically stable solutions for the set of equations corresponding 
to the above set of selfconsistent RPA equations Eqs.(2-6) in 2D as 
well as in 3D. That is a first indication that 
the Fermi liquid properties in the B-F model are destroyed. 

Let us next attempt to understand this behaviour
by closer examining those expressions for the self energies.
Considering to begin with the situation of $E_0 \gg k_BT$ 
for which we obtain
\begin{equation}
\Sigma_F(i\omega_n)=\frac{v^2}{N}\sum_{\bf q}
{{n_F(\varepsilon_{\bf q})}
\over {i\omega_n -E_0 + \varepsilon_{\bf q}}}
\end{equation}
which is independent on the wavevector. 
Introducing the imaginary part of this contribution to the 
retarded self energy i.e., 
$\Gamma(\omega) = -2Im \Sigma_F(\omega)$ we find
\begin{eqnarray}
\Gamma(\omega) & = & 2\pi v^2 \frac{1}{N}\sum_{\bf q} 
n_F(\varepsilon_{\bf q}) 
\delta(\omega -E_0+\varepsilon_{\bf q}) \nonumber \\
               & = & 2\pi v^2 n_F(E_0-\omega) 
\rho_0(E_0 + \mu - \omega)
\end{eqnarray}
where we have taken the limit $i\omega_n \rightarrow \omega + i0$. Let 
us now consider for simplicity a constant model density of states such as 
$\rho_0(\omega)=1/2zt$ for $-zt \leq \omega \leq +zt$. 
Upon Hilbert transforming the expression for $\Gamma(\omega)$ 
we obtain for the corresponing real part of the Fermion self energy
\begin{equation}
R(\omega) = \frac{v^2}{2zt} ln \left|{{\omega-E_0} 
\over {\omega-E_0 - \mu}} \right|
\end{equation} 
The excitation spectrum for the Fermions is then determined by the poles 
of the  Fermion Green's function:
\begin{equation}
G_F({\bf k},\omega_{\bf k}) = \left[\omega_{\bf k} 
- \varepsilon_{\bf k} - R(\omega_{\bf k}) 
+\frac{i}{2}\Gamma(\omega_{\bf k}) \right]^{-1}
\end{equation}
where $\omega_{\bf k}$ denote the complex solutions which determine  
the real 
and imaginary part of the energy spectrum. 
As can be seen from the graphical procedure for determining 
the poles of this Green's function, see Fig.1, 
we generally expect three poles to occur. 
The lowest energy real pole corresponds 
to the bonding state, the highest energy real pole to the anti-bonding 
state and 
the complex pole in the intermediary energy range to the non-bonding state. 
Given 
the fact that $E_0 > 0$, 
since the chemical potential always must lie below the Bosonic level, 
the distribution function $n_{\bf k}$ for the Fermions is enterly 
described by 
the lowest energy solution of Eq.(10) and for which 
$\Gamma(\omega_{\bf k}) \equiv 0$. We thus obtain:
\begin{eqnarray}
n_{\bf k} & = & \int_{-\infty}^{+\infty}d\omega 
\delta(\omega -\varepsilon_{\bf k} - R(\omega))n_F(\omega) \nonumber \\
         & \equiv & Z_{\bf k} n_F(\omega_{\bf k}) =  Z_{\bf k} 
\end{eqnarray}
with $Z_{\bf k} \equiv [1-dR/d\omega|_{\omega_{\bf k}}]^{-1}$ 
representing the spectral weight of 
the lowest energy pole $\omega_{\bf k} \leq 0$ of the 
Fermion Green's function i.e.,
$G_F({\bf k},\omega) =Z_{\bf k}/(\omega- \omega_{\bf k} +i0)$. The 
standard way of determining the Fermi wavevector for Fermi liquids is to 
use the relation for the Fermion distribution function derived from Eq.(10), 
together with the condition $\omega_{\bf k} \leq 0$ which leads to 
$\varepsilon_{{\bf k}_F} + R(0) = 0$. The relations describing the quasi-particle properties at the Fermi surface are then given by:
\begin{eqnarray}
\varepsilon_{{\bf k}_F} & = &\frac{v^2}{2zt} ln \left|{{E_0+\mu} 
\over E_0} \right| 
\nonumber \\
\left. \frac{dR}{d \omega}\right|_0 & = & -\frac{v^2}{2zt} \left( 
\frac{1}{E_0} - \frac{1}{E_0+\mu} 
\right)
\nonumber \\
n_{{\bf k}_F} & = & \left[ 1 + \frac{v^2}{2zt} \left(\frac{1}{E_0} 
- \frac{1}{E_0 + \mu} 
\right) \right]^{-1}
\end{eqnarray}
The B-F model thus predicts two well separated peaks at the Fermi wave vector
whose energies are given by $\omega \simeq 0$ and $\omega \simeq E_0+ \mu +
\mu \frac{E_0+\mu}{E_0}\exp-{(E_0+\mu)/v^2}$. 

From the above Eqs. (11,12) we notice that although in general the Fermionic 
distribution function has a discontinuity at ${\bf k}_F$, 
this discontinuity disappears in 
the limit $E_0 \Rightarrow 0$ (which corresponds to the limit 
$T \Rightarrow T_c (\equiv 0$ in our case of a 1D bare density of states) 
and for which $\varepsilon_{{\bf k}_F} \Rightarrow \infty$ and 
$n_{{\bf k}_F} \Rightarrow 0$. The physical reasons for this breakdown 
of Fermi liquid properties becomes obvious by inspection of the 
contribution proportional to $n_B(E_0)$ in the self energy and which 
for $E_0 \ll k_BT$ 
becomes singular. Let us for that purpose suppose that the Bosons have 
aquired some itinerancy i.e., $E_{\bf q}= E_0+ (\hbar q)^2/2 m_B$ 
where $m_B$ denotes some Boson mass\cite{comment}. 
In the limit $E_0 \Rightarrow 0$ the term 
proportional to $n_B(E_{\bf q})$ becomes singular for $q \Rightarrow 0$. 
Putting 
\(
n_B(E_{\bf q}) = n_B(E_{\bf q}) + \delta_{{\bf q},0}(N n_0 - n_B(E_{\bf q}))
\)
leads to 

\begin{equation}
\Sigma({\bf k},\omega) \simeq v^2{n_0 
\over {\omega +\varepsilon_{\bf k} +i\eta}}
\end{equation}
and ultimately to a contribution to the Fermion Green's function, given by

\begin{equation}
G_F({\bf k},\omega) = {{\omega +\varepsilon_{\bf k}} \over {\omega}^2 
- \varepsilon_{\bf k}^2 - v^2 n_0} 
\end{equation}
This is precisely the BCS Green's function with its two 
poles $\omega_{\bf k} = \pm \gamma_{\bf k} \equiv 
\pm \sqrt{\varepsilon_{\bf k}^2 + v^2 n_0}$ 
showing its  gap in the excitation spectrum given by $\sqrt{v^2 n_0}$ 
and a Fermion distribution function $n_{\bf k}=\frac{1}{2}
[1-\varepsilon_{\bf k}/\gamma_{\bf k}]$ which 
has no longer a discontinuity at ${\bf k}={\bf k}_F$.

\section{Spectroscopic signatures of a two-component scenario}

After these phase space considerations for the spectral 
properties of the B-F model, let us now come to the numerical solution 
of this problem. 
Following the procedure discussed previously\cite{Ranninger-95}
we derive the Fermionic spectral function 
\begin{equation}
A({\bf k},\omega) = - 2 \; Im \; G({\bf k}, \omega+i0)
\end{equation}
at ${\bf k}_F$. There is no unambiguous way of determining  ${\bf k}_F$ 
for an interacting system at finite temperatures. 
We follow here the method spectroscopists use, 
namely to  determine  ${\bf k}_F$ as that vector for which 
the distribution function
\begin{equation}
n_{\bf k}^F = \int_{-\infty}^{+\infty} \frac{d\omega}{2 \pi}  
A({\bf k},\omega) n_F(\omega)
\end{equation}
has an inversion in slope. 
As in our previous calculations we use here as representative parameters 
$v=0.1$  and $\Delta_B = 0.4$ in units of the band width $8t$ 
and a total density of charge carriers $n = 2n_F + 2n_B = 1$. 
The number of Fermions per spin is given by $n_F=\frac{1}{N}\sum_{\bf k} 
n_{\bf k}^F$. 
The values of  ${\bf k}_F$ determined in this way are given as a 
function of temperature in Table I, where we compare them 
with values derived by other criteria such as $n_{\bf k_F}^F = 0.5$ 
and using the Luttinger sum rule $\sum_{k<k_{F}} 1 = Nn_F$.

In Fig.2 we illustrate $A({\bf k}_F,\omega)$ for a set of temperatures 
from which we can see the evolution of the peak structure of this spectral 
function
at the Fermi level. 
We also observe that the lowest energy peak, corresponding to the 
bonding two-particle state, shifts to energies below the Fermi energy 
(equal to zero in our  notation) while the non-bonding single particle state 
shifts to  energies above the Fermi level, 
thus idicating the opening of a pseudogap as the temperature is reduced. 
These findings are in qualitative agreement with the recent experiments 
of ARPES on the underdoped 
high $T_c$ materials\cite{Marshall-96,Ding-96}. 

Our theoretical predictions of the evolution of the three  
pole structure of the Fermionic spectral function
should be a testing ground for

i) the two component scenario of high $T_c$ superconductivity

ii) the pre-existence of bound electron pairs which would be at the origine 
for the deviations from Fermi liquid properties and responsible 
    for the opening of the pseudogap.

ARPES and BIS experiments should be able to see those features,
inspite of the relatively poor resolution available for these type of 
experiments at present. 

The existence of the three peaks in the spectral function can be 
infered from 
the solution of the B-F model in the atomic 
limit for which they are given by $\varepsilon_{\pm}=
(\varepsilon_0 + E_{0}/2) \pm \gamma$ for the lowest energy bonding 
and highest energy anti-bonding two particle states respectively and by 
$\varepsilon_0$ for the intermediary  non-bonding single particle state. 
It should be noticed that those expressions for the two particle poles 
with 
$\gamma=\sqrt {(\varepsilon_0-E_0/2)^{2}+v^2}$ correspond to those  
in the Green's function given by Eq.(14).

The deviations from Landau Fermi liquid behaviour are contained in 
the frequency dependence of the Fermionic self energy. 
We trace the real and imaginary part of this function 
for different temperatures in Fig.(3).
As we can see and as we have pointed out previously, this function does 
not go to zero at the Fermi level when the temperature decreases, 
as would be expected to happen for a standard Fermi liquid. 
We also notice from Fig.(3) 
that the real part of the Fermionic self energy is indeed very similar 
to the model self energy discussed above and which led to the form given 
by Eq.(9). 
This guarantees that the essential physics of the normal state 
properties of the B-F model, 
as discussed above in terms of the model density of states, is basically
correct and should be rather insensitive to dimensionality. 

Our work totally neglects correlation effects as well as such important 
questions as the symmetry of the pseudogap and the incoherent background, 
seen in ARPES and associated with this pseudogap in certain 
direction\cite{Shen-97}. 
Such questions are presently being considered by us and will make 
the topic of some future publication.

\newpage

%\captions

\begin{figure}
{\bf Fig.1} Plot of the real and imaginary part of the Fermion 
self energy for the constant model density of states and the graphic 
solution of the poles for the Fermion Green's function.
\end{figure}

\begin{figure}
{\bf Fig.2} The temperature dependence of the Fermion spectral function
$A({\bf k}_F,\omega)$ 
at the Fermi vector
showing the opening of the pseudogap and the evolution of a three-pole 
structure as the temperature is decreased (a). The temperature evolution 
of the photo emission spectrum intensity (b). $T=0.007$ solid line, $T=0.01$
dashed line and $T=0.02$ dotted line.
\end{figure}

\begin{figure}
{\bf Fig.3} The temperature dependence of the imaginary part (a) and 
the real part (b) of the 
Fermion self energy at the Fermi vector, $T=0.007$ solid line, $T=0.01$
dashed line and $T=0.02$ dotted line.
\end{figure}

\newpage

\begin{table}
\begin{center}
\begin{tabular}{|c|c|c|c|c|} \hline  
\makebox[1.3cm]{\centering $T$}      &
\makebox[2.0cm]{\centering $n_F$} &
\makebox[0.9cm]{\centering ${\bf k}_F (a)$} & 
\makebox[0.9cm]{\centering ${\bf k}_F (b)$} &
\makebox[0.9cm]{\centering ${\bf k}_F (c)$} 
\\  \hline\hline
0.020  &  0.276  & 145 & 143 & 138  \\  \hline
0.015  &  0.230  & 147 & 147 & 146  \\  \hline
0.010  &  0.283  & 150 & 148 & 142  \\  \hline
0.007  &  0.284  & 151 & 149 & 142  \\  \hline
0.006  &  0.285  & 152 & 150 & 143  \\  \hline
\end{tabular}
\end{center}
\caption{The temperature dependence of $n_F$ 
and ${\bf k}_F$ in units of $\pi/1001$. 
We compare ${\bf k}_F$ as determined by 
$d^2n_{{\bf k}}/d{\bf k}^2=0$ (a)  with ${\bf k}_F$ as determined by
$n_{{\bf k}_F}=\frac{1}{2}$ (b) and with ${\bf k}_F$ 
determined by the Luttinger sum rule 
$\sum_{{\bf k} \leq {\bf k}_F} 1 = Nn_F$ (c).} 
\end{table}


\begin{references}
\bibitem{comment} As we have shown in our previous work 
ref(\cite{Ranninger-95,Ranninger-96}), 
as well as in some forthcoming publication of the B-F model 
in infinite dimension,
the dimensionality of the bare density of states does not influence 
the qualitative features of our results.
\bibitem{comment} As shown in our fully conserving treatment of the B-F model, 
refs.\cite{Ranninger-95,Ranninger-96} the intrinsicaly localized Bosons 
aquire itinerancy upon lowering the temperature.
\end{references}
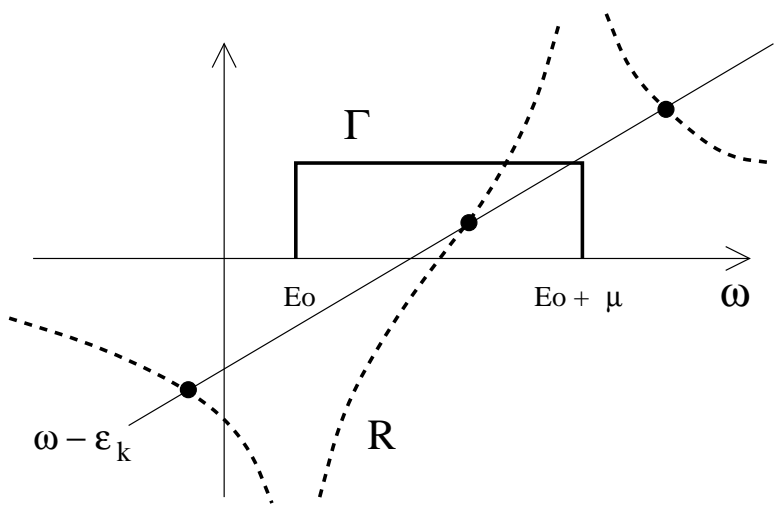
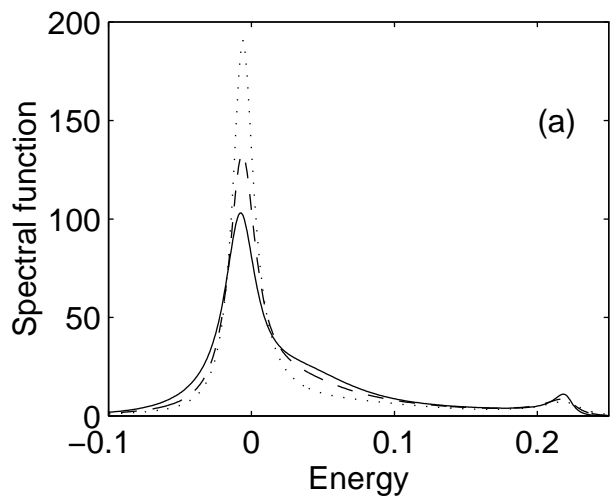
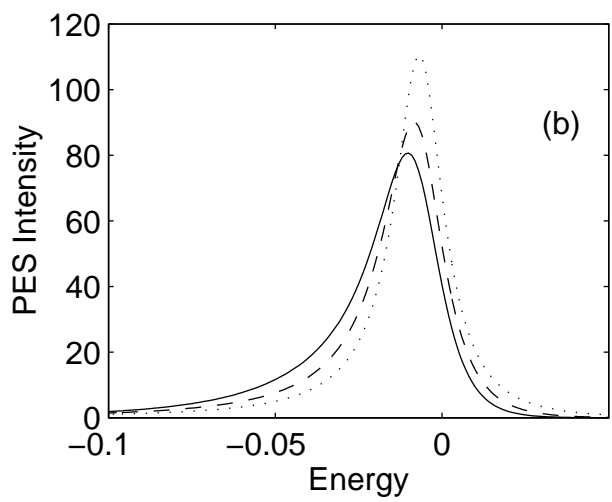
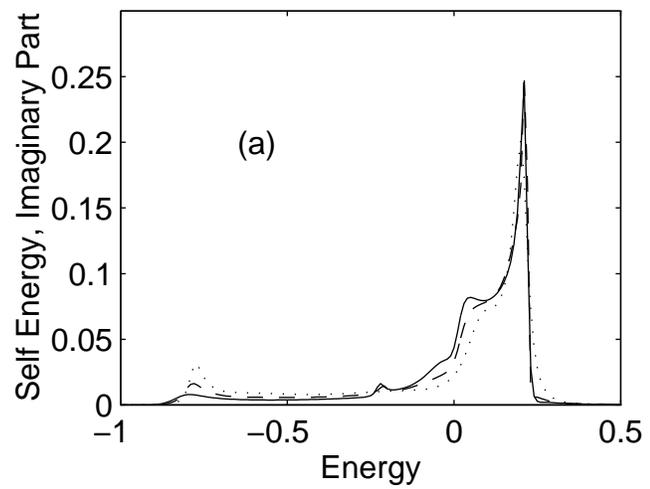
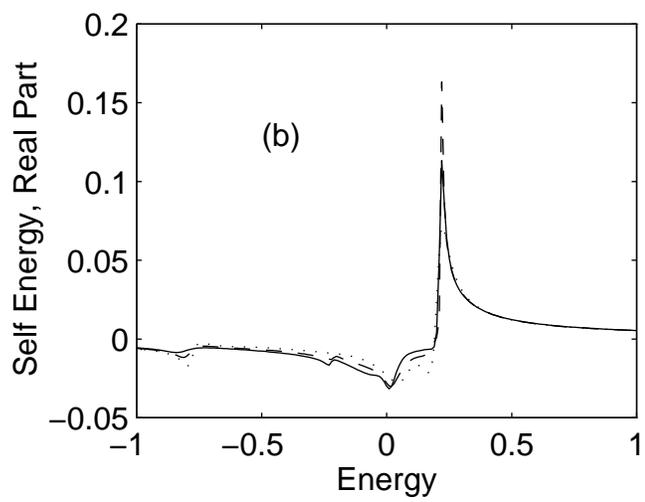
\end{document}